\documentclass[12pt]{article}
\usepackage{amsmath,amsthm,amsfonts,graphicx,epsfig,mathrsfs}
\usepackage[usenames]{color}
\numberwithin{equation}{section}
\newcommand{\pder}[2]{\frac{\partial #1}{\partial #2}}
\newcommand{\braket}[2]{\langle #1 | #2 \rangle}
\newcommand{\der}[2]{\frac{d #1}{d #2 }}

\newcommand{\abs}[1]{\left| #1 \right|}
\newcommand{\e}{\varepsilon}
\def\half{\mbox{$1\over2$}}

\newcommand{\dbar}{\kern-.1em{\raise.8ex\hbox{ -}}\kern-.6em{d}}
 \font\tenrm=cmr10 
\begin{document}
\title{Swimming in curved space \\ or\\ The Baron and the cat }

\author{J.E.~Avron\thanks{We thank Amos Ori for many useful discussions.\ ~ }
and  O.~Kenneth\thanks{Supported in part by
an ISF grant\ }\ \\
\tenrm\!  Department of
Physics, Technion, Haifa 32000, Israel\\
}%
\date{\today}%

 \maketitle

\begin{abstract}
We  study the swimming of non-relativistic deformable bodies in
(empty)  static curved spaces. We focus on the case where the
ambient geometry allows for rigid body motions. In this case the
swimming equations turn out to be geometric. For a small swimmer,
the swimming distance in one stroke is determined by the Riemann
curvature times certain moments of the swimmer.
\end{abstract}

\section{Introduction}
Every street cat can fall on its feet
\cite{video,play,ref:littlejohn,ref:marsden} but only Baron von
Munchausen ever claimed to have lifted himself by applying
self-forces (actually, by pulling on his hair \cite{ref:munch}).
The reason why none else does is because, as we were taught in
high-school physics, the motion of the center of mass can not be
affected by internal forces. In particular, if the center of mass
is at rest, it can  only be moved by external forces.
As pointed out by J. Wisdom \cite{ref:blau,ref:wisdom}, in a
curved space this piece of high school physics is no longer
true: A deformable body can apply internal forces that would
result in swimming, much like a cat turning in Euclidean space
while maintaining zero angular momentum \cite{play}.

\begin{figure}[ht]
\hskip 5 cm\includegraphics[width=4cm]{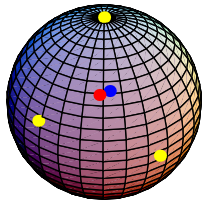} \caption{\em The
red dot is equidistant from the three identical yellow point
masses and so is one natural notion of a center of mass. The blue
dot is another center of mass. It is obtained by first moving the
pair of points masses on the right to their center of mass and
then finding the center of mass of the remaining pair of (unequal)
masses.
 }\label{fig:com}
\end{figure}

The reason why high school physics fails for curved space is tied
to the fact that the notion of center of mass is a fundamentally a
Euclidean notion, with no analog in curved space, see
Fig.~\ref{fig:com}.

Swimming is the motion of deformable bodies.
More precisely, we shall mean by this the rigid body motion affected
by a periodic deformation.
After completing a swimming stroke the swimmer regains its original
shape and finds itself in a different location. The new location
is determined by solving the equations of motion \cite{ref:ws}.

In a generic curved space, the shape of a general swimmer
completely determines its location. Therefore a strictly periodic
deformation allows the swimmer merely to wriggle. To be able to
swim anywhere and with arbitrary orientation the ambient space
must be homogeneous and isotropic. This is the case for the
Euclidean space, and also for symmetric spaces such as spheres and
hyperbolic spaces\footnote{If one thinks of the ambient space in
terms of the underlying space-time structures of general
relativity, then the spaces we shall consider are static Einstein
manifolds \cite{ref:besse}.}. As we shall see, swimming in such
spaces is geometric.

Swimming in more general curved spaces that do not necessarily
allow for rigid body motions require the replacement of shape
space by a different space of controls\footnote{Wisdom takes a
weaker notion of shape which applies to certain tree like
structures that allows for swimming in general space-times.}. Here
we shall focus on the considerably simpler case where the ambient
space
allows 
for rigid body motions.

What do we mean by swimming? Swimming is associated with the
translation bit of a rigid body motion. Recall that even in a
Euclidean space splitting the translations from the rotation for a
given rigid body motion requires picking a fiducial point
\cite{ref:landau}. If we let a turning cat pick the tip of its
tail as a favorite fiducial point it may claim to be Baron von
Munchausen since turning about its center of mass is equivalent to
turning about the tip of the tail accompanied by translation. The
cat is not really swimming because in Euclidean space swimming is
{\em defined} as the the translation of the {\em center of mass}.
Since in a curved space there is no natural notion of a center of
mass we need to specify what does one mean by swimming. We shall
call swimming a motion that after the completion of a stroke,
results in a rigid body motion that does not leave any point
inside the swimmer fixed.

Rigid body motions are generated by the isometries of space. These
are represented by Killing vector fields which we shall denote by
$\xi_{\beta}(x)$. There are $k={d+1\choose 2}$ such fields in $d$
dimensional maximally symmetric spaces. A natural notion of
translations is associated with those Killing fields that reduce
to the Euclidean translations at some fiducial point of the
swimmer. If the swimmer is sufficiently small relative to the
length scale given by the curvature, the Killing fields associated
with these translations will not vanish near the swimmer. This
means that no point of the swimmer is stationary under the rigid
motion generated by such Killing fields and gives a natural notion
of swimming.

As we shall see, a deformable body can generate a rigid body
motion associated to the Killing field $\xi$, if the two-form
$d\xi$ is non-zero near the swimmer. Thus, for example, a cat can
turn in Euclidean space because the Killing two forms associated
to rotations do not vanish. The Baron, in contrast, must be lying
because the Killing two forms associated to translations in
Euclidean space vanish identically. In a curved space any rigid
body motion becomes possible because, in general, none of the
Killing two forms vanish identically. Since the Killing two-forms
associated to translation are proportional to the Riemann
curvature swimming in a curved space is proportional to the
Riemann curvature.

We shall restrict ourselves to studying non-relativistic swimmers
so that the swimming problem reduces to a problem in Newtonian
mechanics on a given static curved space. This is a simplified
version of the problem of the motion of extended bodies in general
relativity which is notoriously difficult \cite{ref:dixon}.

Our two main results are Eq.~(\ref{eq:holonomy}) and
Eq.~(\ref{eq:holonomy-local}). The first gives the swimming
distance of a small swimmer performing an infinitesimal stroke
generated by any pair of deformations. The second is the special
case of linear deformations. In this case, the swimming distance
is proportional to the Riemann curvature and certain moments of
the swimmer. The overall structure of the formulas is similar to
the formula written by Wisdom \cite{ref:wisdom} although both the
setting, the details and the consequences are
different\footnote{For example, Wisdom finds that it is possible
to swim also in those directions that do not admit rigid body
motions while our swimmers obviously can't do that.}.



\section{Swimming from conservation laws}

\subsection{Constants of motion: N\"other theorem}
Consider a body made of a collection of $N$ particles with mass
$m_n$ located at generalized coordinates $x_n$, with $n=1,\dots
N$. Let $L(\dot x, x,t)$ denote the (time-dependent) Lagrangian of
the system which admits a symmetry associated with the Killing
field $\xi (x)$, i.e.  $L$ is invariant under the shift $x\to x+\e
\xi(x) $. The symmetry implies the conservation of
\begin{equation}\label{eq:noether}
P_\xi=\sum_n \left(\pder{L}{\dot x_n^i}\right)\,\xi^i(x_n)\,.
\end{equation}
Indeed the symmetry of the Lagrangian and the Euler-Lagrange
equations of motions imply:
\begin{equation}\label{eq:symmetry}
0=\pder{L}{\e}=\sum_n\left(\pder{L}{\dot x_n}\, \dot \xi(x_n)
+\pder{L}{x_n} \,\xi(x_n) \right)=\der{P_\xi}{t}
\end{equation}
Suppose the dynamics takes place on  manifold with metric $g$. The
(time dependent) Lagrangian, being the difference of the
non-relativistic kinetic energy and the potential energy, is:
\begin{equation}\label{eq:lagrangian-many}
L=\frac 1 2\sum_n m_n\, \dot x_n\cdot\dot x_n-\sum_{pairs}
V_{mn}\big(dis(x_n,x_m),t\big)
\end{equation}
The Lagrangian is  invariant under isometries of the manifold. The
potential energy, being a function of the relative distances,
guarantees Newton third law: The force that particle $a$ applies
on $b$ is directed along the geodesic separating the two and is
opposite and equal to the force that $b$ applies on $a$.

\begin{figure}[ht]
\vskip -1 cm\hskip 4 cm\includegraphics[width=4cm]{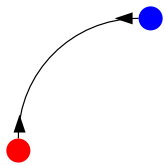}
\caption{\em Newton third law in curved space: The mutual forces
are directed along a geodesic and are balanced.
 }\label{fig:newton}
\end{figure}

The constant of motion $P_\xi$, associated to the isometry $\xi$,
is independent of $V$ and is given by
\begin{equation}\label{eq:momentum}
P_\xi=\sum_n m_n\,\dot x_n\cdot\,\xi(x_n)
\end{equation}
{\em We shall assume throughout that the particles that make up
the body are initially at rest, $\dot x_n=0$, and so $P_\xi=0$.}

Suppose now that the body, initially at rest, can control its
shape by, for example, controlling $V$. The rigid body part of the
resulting motion is determined by the conservations laws
\begin{equation}\label{eq:swimming}
\sum_n m_n\,\xi_{(\beta)}(x_n)\cdot d x_n=0, \quad
\beta=1\dots,k
\end{equation}
It is evident that the resulting rigid body motion is geometric in
the sense that the time parametrization disappeared. 

\section{Deformations and rigid body motions}

We want to describe the (infinitesimal) motion of a deformable
body as a combination of an (infinitesimal) deformation and rigid
body motion. We write the infinitesimal displacement of the n-th
point mass in the form
\begin{equation}\label{eq:first-order}
{dx_n}^j={\xi_{(\beta)}}^j(x_n)\, \dbar\tau^{(\beta)}+{\eta_{(b)}}^j(x_n)\,
d\sigma^{(b)}\,,
\end{equation}
where $\xi_{(\beta)}\, \beta=1,\dots,k $ are Killing fields that
generate rigid body motions and $\eta_{(b)}(x),\ b=1, \dots \ell$
are deformation vector fields that do not. The Killing fields
satisfy the Killing equation \cite{ref:wald}
\begin{equation}\label{eq:killing-deff}
\nabla_{(j}\xi_{k)}=\nabla_j\xi_k+\nabla_k\xi_j=0
\end{equation}
which can be interpreted as the statement that rigid body motion
induces no strain.  The deformation fields, in contrast, induce a
non-vanishing strain
\begin{equation}\label{eq:deformation-deff}
du_{jk}=\big(\nabla_j\,\eta^{(b)}_k+\nabla_k\,\eta^{(b)}_j\big)d\sigma_b
\end{equation}
We may then think of $d\sigma_b$ as the infinitesimal coordinates
of the swimmer's  control space.

Substituting Eq.~(\ref{eq:first-order}) in Eq.~(\ref{eq:swimming})
yields  $k$ linear relations between the infinitesimal rigid body
motion $\dbar\tau$ and the infinitesimal deformations $d\sigma$.
These will turn out to be the swimming equations. However, to
derive these we first need to put coordinates on shapes and rigid
motions that will effectively replace the possibly very large
number of particle coordinates $x_n$.



\subsection{Coordinates for rigid body location}
Let $\tau=0$ denote the initial location of the swimmer, the shape
$S(0)$. We may identify $\tau\in \mathbb{R}^k$ as the new location
of the swimmer (shape) that underwent rigid body motion from the
origin. More precisely, the flow
\begin{equation}
\dot x^j_t=\tau\cdot\xi^j(x_t)=\tau^\alpha\,{\xi_\alpha}^j(x_t)
\end{equation}
at one unit of time gives the displaced shape.  For small $\tau$
the mapping $x\to x_\tau$ is given by
\begin{equation}\label{eq:dx1order}
x^j_\tau=x^j+\tau^\alpha {\xi_\alpha}^j(x)+\half\,
\tau^\alpha\tau^\beta\,{\xi_\alpha}^i\partial_i{\xi_\beta}^j+\dots
\end{equation}
We write for the displaced shape $S(\tau)$, symbolically,
\begin{equation}
S(\tau)=e^{\tau\cdot\xi}S(0)\,.
\end{equation}


\subsection{Shape coordinates}
We now want to put coordinates on shape space. Let $\sigma=0$
denote the undeformed shape $S(0)$. The deformed shape, without
paying attention to its location, will be denote by $S(\sigma )$
with $\sigma\in\mathbb{R}^\ell$ and $\ell$ being the number of
independent strain (deformation) fields at the disposal of the
swimmer. More precisely,  the deformed shape is covariantly
defined by the solution of the flow
\begin{equation}\label{eq:evolution-radial}
\dot x^j_t=\sigma\cdot \eta^j(x_t)
\end{equation}
after a unit time. For small $\sigma$ the mapping $x\to x_\sigma$
is given by
\begin{equation}\label{eq:dx2order}
x_\sigma^j=x^j+\sigma^b{\eta_{(b)}}^j(x)+\frac 1 2
\sigma^b\sigma^c\,{\eta_{(b)}}^k(x)\,\partial_k{\eta_{(c)}}^j(x)+\dots
\end{equation}
We write the corresponding map of the shape, formally,
\begin{equation}
S(\sigma)=e^{\sigma\cdot\eta}S(0)
\end{equation}
Note that this is {\em not} the physical evolution, since the
latter will be accompanied by a rigid motion.
\subsection{Shape and position coordinates}
The shape and location can be parameterized as by a pair
$(\sigma,\tau),\ \tau\in\mathbb{R}^k,\ \sigma\in \mathbb{R}^\ell$.
$\sigma$ is the deformation coordinate and $\tau$ the location
coordinate. Since deformations do not commute with rigid motions
in general, to assign a pair $(\sigma,\tau)$, we need to choose an
order. A natural choice, which reflects the interpretation that
the rigid motion is a consequence of the deformation, is to take
$S(\tau,\sigma)$ to mean\footnote{For the infinitesimal
deformation considered in subsection \ref{jweyfg}, the effect of
this ordering actually turn out to be negligible.}
\begin{equation}\label{eq:shape}
S(\sigma,\tau)=e^{\tau\cdot\xi}\,e^{\sigma\cdot \eta} \, S(0)
\end{equation}
$e^{\sigma\cdot\eta}$  takes $S(0)$ to $S(\sigma,0)$. By
convention $S(\sigma,0)$ has the same location coordinate $\tau$
as the $S(0)$.  $e^{\tau\cdot\xi}$ is the rigid motion by $\tau$
to the correct physical location.

\subsection{Gauge condition}

A swimmer may be able to control its internal stress and thereby
its internal strain, but not directly the deformation fields
$\eta$.  This reflects the simple fact that   there is no
canonical way to integrate Eq.~(\ref{eq:deformation-deff}) to pair
a deformation field with a strain since there is ambiguity in the
Killing fields. To pick a unique $\eta$ one needs to impose $k$
gauge conditions. To do so it is convenient to introduce a notion
of scalar product for extended bodies.

Given a body made of a collection of point particles with masses
$m_n$ at locations $x_n$ we define a scalar product of two vector
fields $u,v$ by:
\begin{equation}\label{eq:scalar-product}
\braket{u}{v}=\frac 1 M\sum_n m_n\,u_j(x_n)\, v^j(x_n),\quad
M=\sum m_n\,.
\end{equation}

We pick the gauge where each deformation field $\eta$ is
orthogonal to all the Killing fields:
\begin{equation}\label{eq:gauge}
\braket{\xi_\alpha}{\eta}=0\,,\quad \alpha=1,\dots, k\,.
\end{equation}
This gauge will play a role in obtaining simple expression for
swimming.


\section{The swimming equations}\label{jhcdg}

\subsection{Covariant description of swimming}\label{jweyfg}


In this subsection we calculate the translation $\Delta\tau$
associated with a small loop $\sigma(t)$ near the origin of shape
space. Applying the deformation forces the particles making up the
body to move along certain trajectories $x_n(t)$. At the leading
order one has by Eqs.~(\ref{eq:dx1order},\ref{eq:dx2order}) simply
\begin{equation}
dx^j_n=\xi^j_{(\alpha)}\,d\tau^{(\alpha)}+
\eta_{(b)}^jd\sigma^{(b)}+O(\sigma d\sigma),\quad
\xi^j_{(\beta)}\big(x_n(t)\big)=\xi^j_{(\beta)}+O(\sigma),
\end{equation}
 and the rhs of both equations is evaluated at $x_n(0)$.
Substituting in the conservation law (\ref{eq:swimming}) we find a
relation between the differential rigid motion, $\dbar\tau$,
and the differential of the strain, $d\sigma$:
\begin{equation}\braket{\xi_{(\beta)}}{\xi_{(\alpha)}}\,\dbar\tau^{(\alpha)}+
\braket{\xi_{(\beta)}}{\eta_{(b)}}\,d\sigma^{(b)}=O(\sigma
d\sigma),\quad \beta=1,...k.\end{equation} The gauge condition
(\ref{eq:gauge}) is now seen to imply (in fact is equivalent to
the statement) that $d\tau=O(\sigma d\sigma)$ i.e. that
$\tau=O(\sigma^2)$. Using this and equations
Eqs.~(\ref{eq:dx1order},\ref{eq:dx2order}) again we see that to
the next order in $\sigma$
\begin{equation}
dx^j_n={\xi_{(\alpha)}}^j\, \dbar\tau^\alpha+ {\eta_{(b)}}^j
d\sigma^b +\frac1 2 \,d(\sigma^b\,\sigma^c)\, \eta_{(c)}^k
\partial_k \eta_{(b)}^j\Big|_{x_n} +O(\sigma^2
d\sigma)\end{equation}
\begin{equation} \xi^j_{(\beta)}\big(x_n(t)\big)= \xi^j_{(\beta)}+
\sigma^c\,{\eta_{(c)}}^k
\partial_k\xi^j_{(\beta)}\Big|_{x_n}+O(\sigma^2)
\end{equation}
Substituting these two equations in the conservation law
(\ref{eq:swimming}) (and again using (\ref{eq:gauge})) we find
\begin{equation}
\braket{\xi_{(\beta)}}{\xi_{(\alpha)}}\,\dbar\tau^\alpha+
\braket{\partial_k{\xi_{(\beta)}}_j}{{\eta^k_{(c)}}\,{\eta^j_{(b)}}}\,
\sigma^c\,d\sigma^b+\frac 1 2
\,\braket{{\xi_{(\beta)}}_j}{\eta_{(c)}^k\,\partial_k{\eta^j_{(b)}}}\,
d(\sigma^b\,\sigma^c)=O(\sigma^2d\sigma)\nonumber
\end{equation}
where the brackets are evaluated for the undeformed body, (and so
are constants, independent of $\sigma$ and $\tau$). The holonomy
follows by applying the $d$ operation. Since $d(\sigma^b\sigma^c)$
is closed, it is annihilated by $d$ and the holonomy
$\delta\tau=d\dbar\tau$ of an infinitesimal cycle of strains is
determined by the linear system of equations:
\begin{equation}
\braket{\xi_{(\beta)}}{\xi_{(\alpha)}}\,\delta\tau^\alpha+
\braket{\partial_k\xi_{(\beta)j}}{\eta_{(c)}^k\,\eta_{(b)}^j}\,
d\sigma^c\wedge d\sigma^b=0\,,\quad \beta=1,\dots,k
\end{equation}
 Using the antisymmetry of forms and the freedom to
relabel indices one can rewrite the second term as
\begin{eqnarray}
 \braket{\partial_k\xi_{(\beta)j}}{\eta_{(b)}^k\,\eta_{(c)}^j}\,
d\sigma^b\wedge d\sigma^c &=&
-\braket{\partial_j\xi_{(\beta)k}}{\eta_{(b)}^k\,\eta_{(c)}^j}\,
d\sigma^b\wedge d\sigma^c\nonumber \\
&=& \frac 1 2
\braket{\partial_{[k}(\xi_\beta)_{j]}}{\eta_{(b)}^k\,\eta_{(c)}^j}\,
d\sigma^b\wedge d\sigma^c\nonumber\\
&=&\frac 1 2 \,\braket{d\xi_\beta}{\eta_{(b)}\,\eta_{(c)}}\,
d\sigma^b\wedge d\sigma^c
\end{eqnarray}
$d\xi$ is the two form associated to the Killing field $\xi$.
Putting all of this together we get the key result, that the
Euclidean motion (the holonomy) is determined by the linear system
of equations:
\begin{equation}\label{eq:holonomy}
\braket{\xi_{(\beta)}}{\xi_{(\alpha)}}\ \delta\tau^\alpha+ \frac 1
2 \, \braket{d\xi_{(\beta)}}{\eta_{(b)},
\eta_{(c)}}\,d\sigma^b\wedge d\sigma^c=0\,,\quad \beta=1,\dots,k
\end{equation}
Where the Killing fields $\xi_{(\alpha)}$ and the deformation
fields, $\eta_{(b)}$, satisfy the gauge condition $\braket{
\xi_{(\alpha)}}{\eta_{(b)}}=0$.



\section{The Killing two form} Let $\xi_{(\alpha)}$ denotes a Killing field
associated with a certain rigid body motion (e.g. translation or
rotation), the question whether one can or can not swim in this
direction depends on purely geometric properties of the ambient
space.   The answer is particularly simple if we
choose the Killing fields to be mutually orthogonal:
$\braket{\xi_{(\beta)}}{\xi_{(\alpha)}}=0$ for $\alpha\neq\beta$.
From  Eq.~(\ref{eq:holonomy}) we see that a rigid motion generated
by $\xi_{(\alpha)}$ is possible only if the corresponding two-form
$d\xi_{(\alpha)}$ does not vanish identically on shape space.

\subsection{Killing fields in the Euclidean space}

With $x^j$ Cartesian coordinate in a Euclidean space, the Killing
(vector) fields generating translation in the k-th direction is
\begin{equation}\label{eq:killing-euclid}
\xi_{(k)}=\partial_{k}
\end{equation}
The corresponding one forms, $ \xi_{(k)}=d{x^k}$ are evidently
exact.

The Killing fields generating rotations in the $j-k$ plane about
the origin are
\begin{equation}\label{eq:killing-rot}
\xi_{(jk)}=x^j\partial_{k}-x^k\partial_{j}
\end{equation}
The corresponding one-forms are
$\xi_{(jk)}=x^j\,d{x^k}-x^k\,d{x^j}$.  These are not closed:
\begin{equation}\label{eq:2form}
d\xi_{(jk)}=2dx^j\wedge d{x^k}
\end{equation}

 In Euclidean space the Killing fields of orthogonal
translations  are evidently orthogonal
$\braket{\xi_j}{\xi_k}=0$ for $j\neq k$. A trite calculation shows
that the Killing fields corresponding to translations and
rotations are mutually orthogonal, $\braket{\xi_j}{\xi_{kl}}=0$,
provided the rotations are about the center of mass. Similarly,
the rotations about the principal axis of a body are mutually
orthogonal.

Since $d\xi_{(k)}=0$ for the translations the Baron must be lying.
Cats, in contrast, can turn in Euclidean space because of the
non-vanishing of $d\xi_{(jk)}$.

\subsection{Local Euclidean frames}\label{sec:local-euclid}
By a local Euclidean coordinate system we mean a coordinate system
where the metric tensor is the identity at the origin,
$g_{ij}(0)=\delta_{ij}$ and the Christoffel symbols vanish there
$\Gamma^i_{jk}(0)=0$. Such a coordinate systems always exists. (It
is not unique, however, as $g_{ij}(0),\Gamma^i_{jk}(0)$ are
unaffected by any coordinate transformation of the form
$x'^i=x^i+O(x^3)$, as well as by orthogonal transformations.) A
local Euclidean system allows us to extend Euclidean notions, such
as translations, linear deformations and a center of mass, to the
slightly non-Euclidean setting.
In local Euclidean coordinates, relations which hold in Euclidean geometry
will typically get corrections of relative order $O(Rx^2)$
where $R$ is the space curvature.

\subsection{Moments}
For a  small swimmer we define the multipole moments in a local
Euclidean coordinates by
\begin{equation}\label{Q}
Q^{ij\dots k}=\sum m_n\, x_n^ix_n^j\dots x_n^k
\end{equation}
The moments are well defined\footnote{ In general moments are
coordinate dependent and hence non-covariant.} up to terms which
are smaller by order $O(R\,L^2)$ where $L$ is the linear dimension
of the swimmer. In particular, the vanishing of the first moments
$Q^j=0$ says that the (approximate notion of) the center of mass
is at the origin. This is a natural choice of the origin which we
shall normally make.

\subsection{The translation Killing two form in curved
spaces}\label{ch:killing-curved}

Swimming in a curved space is possible because the Killing fields
corresponding to translations are not closed $d\xi_{(k)}\neq 0$.
Here we shall explain what we mean by the Killing fields
associated to translations in a curved space and calculate the
leading behavior of the corresponding two form.

Let $x^j$ be a locally Euclidean coordinate system in the
neighborhood of the origin. A Killing field $\xi$, if it exists,
is uniquely determined by its value at the origin, $\xi_i(0)$, and
its (necessarily antisymmetric) first derivative
$\nabla_{[i}\xi_{j]}(0)$ there\footnote{Even when an actual
Killing field does not exit, this initial data is enough to
determine an approximate Killing field near the origin.}. It is
natural then to associate with translation along the $k$ axis the
Killing field $\xi_{(k)}$ which satisfies
\begin{equation}\label{eq:leadingoderkilling}
\xi_{(k)i}(0)=\delta_{ki},\;\nabla_{[i}\xi_{(k)j]}(0)=0
\end{equation}
Its leading behavior away from the origin can be determined by the
Riemann curvature through the differential equation
\cite{ref:wald}
\begin{equation}
\nabla_i\nabla_j\xi_l=-{R_{jli}}^h\xi_h
\end{equation}
Since $\Gamma(0)=0$, at the origin
$\nabla_i\nabla_j\xi_l=\partial_i\nabla_j\xi_l$, substituting the
initial data Eq.~(\ref{eq:leadingoderkilling}) on the right hand
side and integrating, give
\begin{equation}\label{eq:approximate2form}
\nabla_j(\xi_{(k)})_\ell=- x^i R_{j l i k}(0)+O(x^2)
\end{equation}
These are the $j-l$ components of the k-th Killing two form. These
are evidently anti-symmetric in $j,l$, by the properties of the
Riemann tensor. This guarantees the Killing equation
$\nabla_{(i}\xi_{j)}=0$ to the leading order. In appendix
\ref{ap:constant-curvature} we give exact and explicit examples
of such Killing fields and their two forms.

\section{Small swimmers}\label{ch:riemann}
The swimming equations take a more transparent form in a
coordinate system which is approximately Euclidean near the
swimmer. This requires that the swimmer be small. This is the case
if its linear dimension $L$ is such that $RL^2\ll 1$, where $R$ is
the curvature. Swimming then corresponds to the Killing fields
associated with the translations described in subsection
\ref{ch:killing-curved}.


Let $x^j$ be a locally Euclidean coordinate system in the
neighborhood of the swimmer. We choose the origin so the
(approximate) center of mass is at the origin: $Q^j=0$. This
guarantees that (to leading order in $L^2 R$) the Killing fields
associated to rotations and translations are mutually orthogonal
and one can therefore study the translations independent of the
rotations.

Let $\eta_{(b)}$ and $\eta_{(c)}$ be two deformation field
satisfying the gauge conditions. An infinitesimal loop in strain
space with area $d\sigma^b\wedge d\sigma^c$, will lead to swimming
a distance $\delta x^k$ along the $k$-axis of a local Euclidean
frame. Combining
Eqs.~(\ref{eq:approximate2form},\ref{eq:holonomy}) gives our main
result\footnote{If $\xi$ is only an approximate Killing field then
momentum is not strictly conserved and one gets an error term for
$\dot P_\xi$ which is of the order $M\,L^4\omega^2 R'\xi dt$ where
$L$ is the linear scale of the swimmer and $\omega$ the frequency
of the stroke. After $N$ strokes this will lead to an error in the
location of the swimmer that is of the order $L^4R'\xi N^2$.   }:
\begin{equation}\label{eq:holonomy-local}
M\delta
x^k=R_{jlik}\,\left(\sum_n\,m_n\,x^i_n\,\eta_{(b)}^j\eta_{(c)}^l
\Big|_{x_n}\right)\,d\sigma^{b}\wedge d\sigma^{c}\,,\end{equation}
Where $R_{jlik}$ is the Riemann curvature at the origin, is a
property of the ambient space while the brackets are a property of
the strained swimmer.

\section{Swimming with constant strains}

In a Euclidean space the $d+1 \choose 2$ linear deformations
\begin{equation}\label{eq:deformation-linear}
2\,\eta_{(jk)}=x^j\partial_k+x^k\partial_j\
\end{equation}
generate constant strains (and thus are evidently transversal to
the rigid body motions). The notion of linear deformations does
not have a covariant meaning in a general curved space. However,
there is an approximate notion of linear deformation associated to
a local Euclidean frame which one can safely apply to small
swimmers.

For a general swimmer, the linear deformations of
Eq.~(\ref{eq:deformation-linear}) will fail to satisfy the gauge
condition, Eq.~(\ref{eq:gauge}). But, the deformation fields can
be tweaked to do so, by adding Killing fields. Explicitly, suppose
that that swimmer has its center of mass as the origin, $Q^j=0$,
and pick the orientations of the Euclidean frame to be the
principal frame of the swimmer with $Q^{jk}$ diagonal. The linear
deformations $\eta_{(jk)}$ defined by
\begin{equation}\label{eq:deformation-ort}
(Q^{kk}+Q^{jj})\,\eta_{(jk)}=x^j\,Q^{kk}\,\partial_k+x^k\,Q^{jj}\,\partial_j
\end{equation}
(no summation over $j,k$), satisfy the gauge condition and induce
(approximately) constant strains. Shape space may then be
identified with $\mathbb{R}^\ell$ where $\ell={d+1\choose 2}$.


For linear deformations the bracket in
Eq.~(\ref{eq:holonomy-local}) is clearly proportional to
$Q^{imh}$, the tensor of cubic moments of the swimmer. Using the
explicit form Eq.~(\ref{eq:deformation-ort}) for the deformation
fields one may write
\begin{equation}\label{eq:holonomy-local1}
M\delta x^k=R_{jlik}\,Q^{imh}\,C^{jl}_{mk}(b,c)
\,d\sigma^{b}\wedge d\sigma^{c}\,,
\end{equation}
The matrix $C^{jl}_{mk}(b,c)$ is non-zero only when the four
outside indices $(j,l,m,k)$ agree pairwise with the four internal
indices $(b,c)=(\beta\beta',\gamma\gamma')$. Explicitly it is
given by
\begin{eqnarray}
C^{jl}_{mk}(\beta\beta',\gamma\gamma')=
{Q^{jj}Q^{ll}\over(Q^{\beta\beta}+Q^{\beta'\beta'})
(Q^{\gamma\gamma}+Q^{\gamma'\gamma'})}\times\hskip 3 cm\\
\big(\delta_{m\beta}\delta_{j\beta'}
+\delta_{j\beta}\delta_{m\beta'}\big)\,\big(\delta_{h\gamma}
\delta_{l\gamma'}+\delta_{l\gamma}\delta_{h\gamma'}\big).\nonumber
\end{eqnarray}
The formula disentangles the space from the swimmer in the sense
that $R_{khij}$ is only a property of space, while the second and
third rank tensors $ Q^{\beta\beta}\,,Q^{h\ell m}$ are properties
of the swimmer. It follows that:

\begin{itemize}
\item The swimming distance scales like the cubic moment of the
swimmer, as it must be by dimensional analysis. (Since the
swimming is proportional to the curvature, and $d\sigma$ are
dimensionless  \cite{ref:wisdom}.) This implies that small
swimmers are heavily penalized.
 \item A body which is invariant under inversion $x^j\to
-x^j$ for all $j$, can not swim via linear deformations since
$Q^{h\ell m}=0$. \item A needle can not swim under linear
deformations and neither can a swimmer made of two point
particles. This is because, by Eq.~(\ref{eq:deformation-ort}), a
single deformation, $\eta_{11}$, (with $1$ the axis of the needle)
satisfies the gauge condition, and at least two are needed to
swim.
\item At least three point particles are required to swim by
linear deformations (as is also evident by other considerations).
\end{itemize}






\appendix
\section{Topological swimmers}\label{ap:picturebook}

\subsection{Swimming on a ring}
\begin{figure}[bht]
\hskip 5 cm\includegraphics[width=4cm]{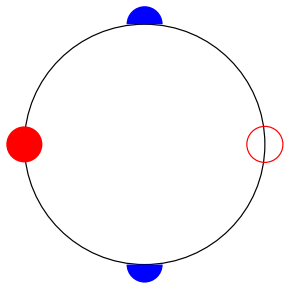} \caption{\em The
full red circle on the left, splits into the two blue half balls
which recombine at the empty red circle on the right. This allows
for moving without violating Newton laws.
 }\label{fig:c-swimmer}
\end{figure}

One can swim even in flat space if the topology is nontrivial. A
ring, being one dimensional, has no curvature. Nevertheless, a
composite body at rest can displace itself by disintegrate into a
pair of particles and recombining as shown in the Figure
\ref{fig:c-swimmer}. The energy needed for breakup is recovered at
the fusion.  The net displacement is determined by the mass ratio
of the two splinters. Ideally, the displacement does not require
dissipation.   The ability to swim in such (flat)
spaces can, once again, be traced to the absence of a good notion
of center of mass.

\subsection{Intersecting geodesics}
\begin{figure}[h]
\vskip -1 cm\hskip 5 cm\includegraphics[width=4cm]{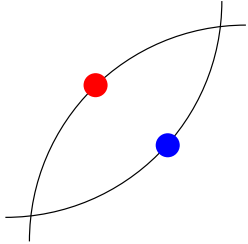}
\caption{\em The red ball can displace itself to the position of
the blue ball by splitting into two parts that it sends to
crossing points of geodesics and then recombine satisfying
Newton's law in the process.
 }\label{fig:intersecting}
\end{figure}
A variation on this theme occurs whenever the space has
intersecting geodesics, as shown in Fig. \ref{fig:intersecting}.
This mode of swimming is non-local since geodesics that start at a
given point do not intersect in a small neighborhood.

\subsection{Converting energy to velocity}
\begin{figure}[hbt]
\hskip 5 cm\includegraphics[height=3.5cm]{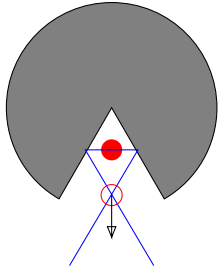} \caption{\em
A cone is represented by the unshaded  sector of the plane. The
straight lines on the boundary of the gray area are identified.
Geodesics are ordinary straight lines in the unshaded area. The
stationary red ball splits and then recombines to a moving (empty)
ball.}\label{fig:cone-swimmer}
\end{figure}

A single particle at rest can disintegrate into a pair and
recombine at a different point in such a way that the recombined
particle has net velocity. Parts of the the energy needed to
disintegrate is not recovered after fusion and remains as kinetic
energy. This is a kinematic process that requires only contact
forces and can take place even on a cone, which is everywhere flat
(except at the apex)  as shown in
Fig.~\ref{fig:cone-swimmer}. The process described swimming in
momentum space.

\section{Surfaces with constant
curvature}\label{ap:constant-curvature}
\subsection{Killing fields}
Surfaces with constant curvature are homogeneous and isotropic and
are locally isometric either to the Euclidean space, ($R=0$), the
sphere ($R>0$), or the hyperbolic space ($R<0$). All can be
stereographically projected to the complex plane with coordinate
$z=x+iy$.
 We choose to normalize $z$ such that near the origin it
reduces to the Euclidean coordinate. The metric is then:
\begin{equation}
(d\ell)^2= \frac{|dz|^2}{(1+R|z|^2)^2}
\end{equation}
$R$ is the Gaussian curvature. The (normalized) Killing field are
then
\begin{eqnarray}\label{eq:killing-sphere-real}
\xi_{1}&=&\partial_x+R\left(
(x^2-y^2)\,\partial_x+2xy\partial_y\right)\,,\\
\xi_{2}&=&\partial_y+R\left(
2xy\,\partial_x+(y^2-x^2)\,\partial_y\right),\nonumber \\
\xi_{3}&=&x\partial_y-y\partial_x\nonumber
\end{eqnarray}

\subsection{Swimming on a surfaces with constant curvature}

Consider swimming along the x-axis of the plane with the metric of
the sphere or the hyperbolic plane. The swimmer is assumed to be
symmetric under reflection $y\to-y $ and to be localized near the
origin. The Killing field $\xi_1$ of
Eq.~(\ref{eq:killing-sphere-real}), generates a flow that preserve
the reflection symmetry and (due to this symmetry) is
orthogonal to the other two Killing fields
\begin{equation}
\braket{\xi_{1}}{\xi_{2}}=\braket{\xi_{1}}{\xi_{3}}=0
\end{equation}
This implies, (by Eq.~(\ref{eq:holonomy})), that the motion
generated by $\xi_1$ decouples from the other rigid motions. For a
small swimmer, {$\abs{Rz^2}< 1$}, $\xi_1$ leaves no point of the
swimmer fixed, since only $\xi_1(\pm i/\sqrt R)=0$,  and describes
a bona fide swimming. The swimming is along the x-axis, by
symmetry. We henceforth write $\xi$ for $\xi_{1}$.

The Killing one-form is
\begin{equation}\label{eq:killing-one-form}
2\,\mathbf{\xi}=\frac{ 1+R z^2}{(1+R\abs{z}^2)^2}\,d\bar z+
\frac{ 1+R \bar z^2} {(1+R\abs{z}^2)^2}\,d z\,.
\end{equation}
The two form is then
\begin{equation}\label{eq:killing-two-form}
d\mathbf{\xi}= 2R\,\frac{ z-\bar z} {(1+R\abs{z}^2)^3}\,dz\wedge
d\bar z\,,
\end{equation}
Immediate consequences of this expression for the two form are:
\begin{enumerate}
\item Since $d\mathbf{\xi}$ is not identically zero one can
generate $\xi$ with appropriate deformation fields. \item Since
$d\xi$ changes sign with $R$,  a swimming stroke that swims to the
right on the sphere swims to the left on the hyperbolic plane.
 \item A deformable one-dimensional body
can not swim in the direction of its axis (under any deformation,
not necessarily linear).
\end{enumerate}

\subsubsection{Small swimmers}

Consider a small swimmer, $RL^2\ll 1$, which is located initially
near the origin in the complex plane. Suppose, as before, that the
swimmer is symmetric under reflection $y\to -y$ and consider
deformation fields preserving the reflection symmetry. This
implies no rotation and the only possible motion is swimming along
the x-axis. For the sake of simplicity suppose that the origin is
chosen so that the swimmer is balanced in the sense that initially
$\sum m_nx_n=0$.

For the motion along the x-axis we have, from
Eqs.~(\ref{eq:killing-sphere-real}, and
Eq.~(\ref{eq:killing-two-form}),
\begin{equation}
\xi= \,\partial_x+O(RL^2\partial)\,,\quad d\mathbf{\xi}= 8R\, y\,
dx\wedge dy+O(R^2L^3)\, dx\wedge dy
\end{equation}
A pair of linear deformation that preserve the reflection symmetry
and satisfies the gauge condition is
\begin{equation}
\eta= by\,\partial_y+cx\partial_x\,, \quad \eta'= b'y\,\partial_y+c'x\partial_x
\end{equation}
For such pair
\begin{equation}
\braket{d\xi}{\eta,\eta'}=- 8R\frac {bc'-b'c} M\sum_n m_n y^2_nx_n
\end{equation}
The swimming distance $\delta x$ (along the x-axis) covered by the
pair of infinitesimal strokes $(\eta\,d\sigma,\eta'\,d\sigma ')$
is,
\begin{equation}
 \delta x \approx- \,\frac {8R} M\,\left(\sum m_n \,
x_n\,y^2_n\right)\,dA
\end{equation}
$dA=(bc'-b'c)d\sigma\wedge d\sigma'$ is the area form in the space of
controls (strains).

\begin{figure}[bht]
\hskip 4 cm\includegraphics[width=4cm]{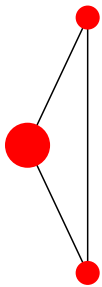}
\caption{\em A swimming triangle on the sphere and the
pseudosphere. The oars are long and their total weight equals the
weight of the payload.
 }\label{fig:triangle}
\end{figure}
\subsubsection{Swimming triangles}
Consider an isosceles triangle made of two identical point masses
$m$ at the base so that the total mass of the swimmer is $M$. For
a the triangle of height $h$ and base $b$ (whose center of mass is
at the origin) one has
\begin{equation}
\frac {8} M\,\left(\sum m_n \, x_n\,y^2_n\right)=4 \frac
{m(M-2m)}{ M^2}\,
 h\,b^2\le \frac 1 2\, h b^2
\end{equation}
and the optimizer is when $4m=M$. The optimal weight distribution
has oars whose weight balances the weight of the payload. The
dependence on $b$ may be interpreted as the statement that a good
swimmer needs long oars. If the swimming stroke is a rectangle in
the $(b,h)$ plane with sides  $(\delta b,\delta h)$ then
$dA=(\delta b/b)(\delta h/h)$ and the swimming distance $\delta x$
is at most $\half b\,\delta b\,\delta h$.






\end{document}